\documentclass[10pt]{elsarticle}
\usepackage{amsmath,amssymb,textcomp,float}
\usepackage{fullpage}
\begin{document}
\title{Operational characteristics of single particle heat engines and refrigerators with time asymmetric protocol.}
\author[mymainaddress]{P. S. Pal \corref{mycorrespondingauthor}}
\cortext[mycorrespondingauthor]{Corresponding author}
\ead{priyo@iopb.res.in}
\author[mymainaddress]{Arnab Saha}
\ead{sahaarn@gmail.com}
\author[mymainaddress]{A. M. Jayannavar}
\ead{jayan@iopb.res.in}
\address[mymainaddress]{Institute of Physics, Sachivalaya Marg, Bhubaneswar - 751005, India}
\begin{abstract}
We have studied the single particle heat engine and refrigerator driven by time asymmetric protocol of finite duration. Our system consists of a particle in a harmonic trap with time-periodic strength that drives the particle cyclically between two baths. Each cycle consists of two isothermal steps at different temperatures and two adiabatic steps connecting them. The system works in irreversible mode of operation even in the quasistatic regime. This is indicated by finite entropy production even in the large cycle time limit.  Consequently, Carnot efficiency for heat engine or Carnot Co-efficient of performance (COP) for refrigerators are not achievable. We further analysed the phase diagram of heat engines and refrigerators. They are sensitive to time-asymmetry of the protocol. Phase diagram shows several interesting features, often counterintuitive. The distribution of stochastic efficiency and COP is broad and exhibits power law tails. 
\end{abstract}
\newcommand{\nwc}{\newcommand}
\nwc{\vs}{\vspace}
\nwc{\hs}{\hspace}
\nwc{\la}{\langle}
\nwc{\ra}{\rangle}
\nwc{\lw}{\linewidth}
\nwc{\nn}{\nonumber}

\nwc{\pd}[2]{\frac{\partial #1}{\partial #2}}
\nwc{\zprl}[3]{Phys. Rev. Lett. ~{\bf #1},~#2~(#3)}
\nwc{\zpre}[3]{Phys. Rev. E ~{\bf #1},~#2~(#3)}
\nwc{\zpra}[3]{Phys. Rev. A ~{\bf #1},~#2~(#3)}
\nwc{\zjsm}[3]{J. Stat. Mech. ~{\bf #1},~#2~(#3)}
\nwc{\zepjb}[3]{Eur. Phys. J. B ~{\bf #1},~#2~(#3)}
\nwc{\zrmp}[3]{Rev. Mod. Phys. ~{\bf #1},~#2~(#3)}
\nwc{\zepl}[3]{Europhys. Lett. ~{\bf #1},~#2~(#3)}
\nwc{\zjsp}[3]{J. Stat. Phys. ~{\bf #1},~#2~(#3)}
\nwc{\zptps}[3]{Prog. Theor. Phys. Suppl. ~{\bf #1},~#2~(#3)}
\nwc{\zpt}[3]{Physics Today ~{\bf #1},~#2~(#3)}
\nwc{\zap}[3]{Adv. Phys. ~{\bf #1},~#2~(#3)}
\nwc{\zjpcm}[3]{J. Phys. Condens. Matter ~{\bf #1},~#2~(#3)}
\nwc{\zjpa}[3]{J. Phys. A  ~{\bf #1},~#2~(#3)}
\nwc{\znjp}[3]{New. J Phys. ~{\bf #1},~#2~(#3)}
\nwc{\zapl}[3]{Applied Phys. Lett. ~{\bf #1},~#2~(#3)}

\maketitle

In recent years a lot of interest has been generated in the study of stochastic single particle heat engines and refrigerators \cite{bro06,naka06,mara07,ron08,sch08,eng13,tu13,hol14,luc15,rana14,rana16}. Engines at nanoscale are ubiquitous in biology \cite{deba,yoshida,marisa} and become increasingly pertinent synthetically. With the progress of technology micrometer  sized stochastic heat engines have been realised experimentally \cite{nphy12,mar14,lutz115,lutz215}. At these length scales thermal fluctuation plays a pivotal role in determining the performance characteristics of the system. Typical energy transformations (work and heat) in these systems are of the order of $k_BT$, where $T$ is the temperature of the surrounding reservoir.Therefore taking account of thermal fluctuations is an absolute necessity to achieve engineering capabilities in designing such small scale devices \cite{rit05}. 

The apt theory for the thermodynamics of small scale devices comes under the frame work of stochastic thermodynamics where the macroscopic thermodynamic variables (e.g. work, heat, total entropy, internal energy etc.) are defined over a single trajectory and thereby differs stochastically from one measurement to another \cite{sek98, sei05, dan05, sai07, jop08, arnab09, sek-book, sai11}. Besides validating macro thermodynamics after averaging over all possible trajectories, the new frame work offers {\it {first law like}} equality defined over a single trajectory and fluctuation theorems \cite{rit03, harris07, jar10, sei12, lah14, lah141}, a set of equalities between stochastically varying thermodynamic variables that put rigorous constraints to their distributions.  

Using stochastic thermodynamics microscopic heat engines and refrigerators have been explored. Extensive studies including both quasistatic and nonquasistatic regime have been done on systems consisting of a harmonically trapped Brownian particle driven periodically (with period $\tau$) by the time dependent strength of the confining potential within two thermal reservers having different temperatures $T_h$ and $T_l$ where $T_h>T_l$  \cite{rana14, rana16}. The protocol studied in \cite{rana14, rana16} consists of two isotherms having equal length along time axis (that is why the protocol can be termed as time-symmetric) and two adiabatic path connecting them by instantaneous jumps. We found that in this case the system operates in four thermodynamically favourable modes: a.) Engine - heat from hot bath is converted partially into work and the rest is supplied to the cold bath, b.) Heater I - work done on the system is divided into two parts that heat up both the baths, c.) Heater II - the system takes heat from hot bath and with the help work done on it, the heat is transferred from hot to the cold bath, d.) Refrigerator -  the system takes heat from the cold bath and with the help of work done transfers heat to the hot bath. Different operational mode of the system appears at different values of $\tau$ and $\frac{T_h}{T_l}$, which are described by the {\it{phase diagram}} in $\tau - T_h$ plane for fixed $T_l$. 


In the following we extend our previous studies \cite{rana14, rana16} by driving the single particle heat engine and refrigerator with time-asymmetric protocols. Here we will analyse the thermodynamics of the system in quasistatic as well as nonquasistatic regime, driven by the protocol having fixed $\tau$ but with {\it unequal} lengths of the isothermal steps along the time axis (i.e. the protocol is time-asymmetric) together with the {\it equal} jumps of the protocol along the adiabatic steps. We find that in nonquasistatic limit, tuning the lengths of the isothermal steps keeping $\tau$ fixed, the phase diagram can be modified. Secondly, we find that in quasistatic regime with high friction, the heights of the adiabatic jumps of the protocol and the ratio of the bath temperatures together will determine a generic condition for the system operating under reversible mode of operation.  




In this paper, first we describe the model and the protocols for the drive. Then we analyse the quasistatic behaviour of the system driven by all the protocols both in the underdamped and overdamped limit. Next, after briefly discussing the basics of stochastic thermodynamics for completeness, we provide detailed analysis of non-quasistatic  behaviour of the system focussing on the effects of time asymmetry of the protocols used. Finally we summarise our result and conclude.

\section{Model}
Our system consists of a Brownian particle of mass $m$ having position $x$ and velocity $v$, confined in a harmonic trap. The stiffness of the trap $k(t)$ is varied periodically in time using a time-asymmetric protocols.  For the underdamped case, the equation of motion of the particle in contact with a heat bath at temperature $T$ is given by \cite{risk-book, coffey-book}
\begin{equation}
m\dot{v}=-\gamma v-k(t)x+\sqrt{\gamma k_BT}\xi(t)
\label{eom1}
\end{equation}
In overdamped limit, the equation of motion  reduces to 
\begin{equation}
\gamma\dot x=-k(t)x+\sqrt{\gamma k_BT}\xi(t).
\label{eom2}
\end{equation}
Here the fluctuation dissipation relation between noise strength, temperature of the bath ($T$) and friction coefficient ($\gamma$) is maintained. In further analysis, the mass of the particle, friction coefficient and  Boltzmann constant $k_B$ are set to unity. The noises from the bath $\xi$ are  Gaussian distributed with zero mean  and are delta correlated, i.e., $\la \xi(t)\ra=0 $ and $\la \xi(t_1)\xi(t_2)\ra=2\delta(t_1 -t_2)$.

Two types of time-asymmetric periodic protocols of periodicity $\tau$ have been applied on our system viz. engine protocol and refrigerating protocol(reverse of the previous one).  Each of the protocol consists of four steps: two isothermal and two adiabatic. The isothermal processes takes place in finite time whereas the adiabatic processes occur instantaneously. During one isothermal process the system is connected to a hot bath at temperature $T_h$ and in the other isothermal process the system is connected to cold bath at temperature  $T_l$. Time-asymmetry in protocols refers to the fact that the contact time with two heat baths during the isothermal processes is different. These protocols are described below.

\underline{Engine protocol}: In the first step the system undergoes an isothermal expansion in contact with hot bath and the stiffness is changed from  an initial value $a$ to $a/2$. In second step the stiffness is changed from $a/2$ to $a/4$ instantaneously to perform the adiabatic expansion process. In this step the system is disconnected from the hot bath and instantly connected to the cold bath. Next, another isothermal process takes place in which the trapped is compressed and the stiffness is changed from $a/4$ to $3a/4$. In the last step adiabatic compression takes place and the stiffness is changed instantaneously from $3a/4$ to $a$. In this step the system is again connected back to the hot bath. The ratio of the contact times with two heat baths during the isothermal processes is $r:s$ i.e., the duration of isothermal expansion is $\tau_1=\left(\frac{r}{r+s}\right)\tau$ and the duration of the isothermal compression is $\tau_2=\left(\frac{s}{r+s}\right)\tau$. The time dependency of stiffness is given in the following equations,
\begin{eqnarray}
 k(t)&=&a\left(1-\frac{r+s}{2r}\frac{t}{\tau}\right) \phantom x\phantom x \phantom x \phantom x\phantom x\phantom x\phantom x 0\leq t < \tau_1  \nonumber\\
 &=&a/4 \phantom x\phantom x \phantom x \phantom x \phantom x\phantom x \phantom x \phantom x  \phantom x\phantom x\phantom x \phantom x \phantom x\phantom x \phantom x \phantom x  \phantom x \phantom x t = \tau_1\nn\\ 
 &=&a\left(\frac{1}{4}-\frac{r}{2s}+\frac{r+s}{2s}\frac{t}{\tau}\right) \phantom x\phantom x \tau_1\leq t < \tau  \nonumber \\
 &=&a \phantom x\phantom x \phantom x \phantom x \phantom x\phantom x \phantom x \phantom x \phantom x \phantom x\phantom x \phantom x \phantom x \phantom x \phantom x \phantom x  \phantom x \phantom x\phantom x\phantom x t = \tau .
 \label{protocol1}
\end{eqnarray}

\begin{figure}[H]
\begin{center}
 \includegraphics[width=10cm]{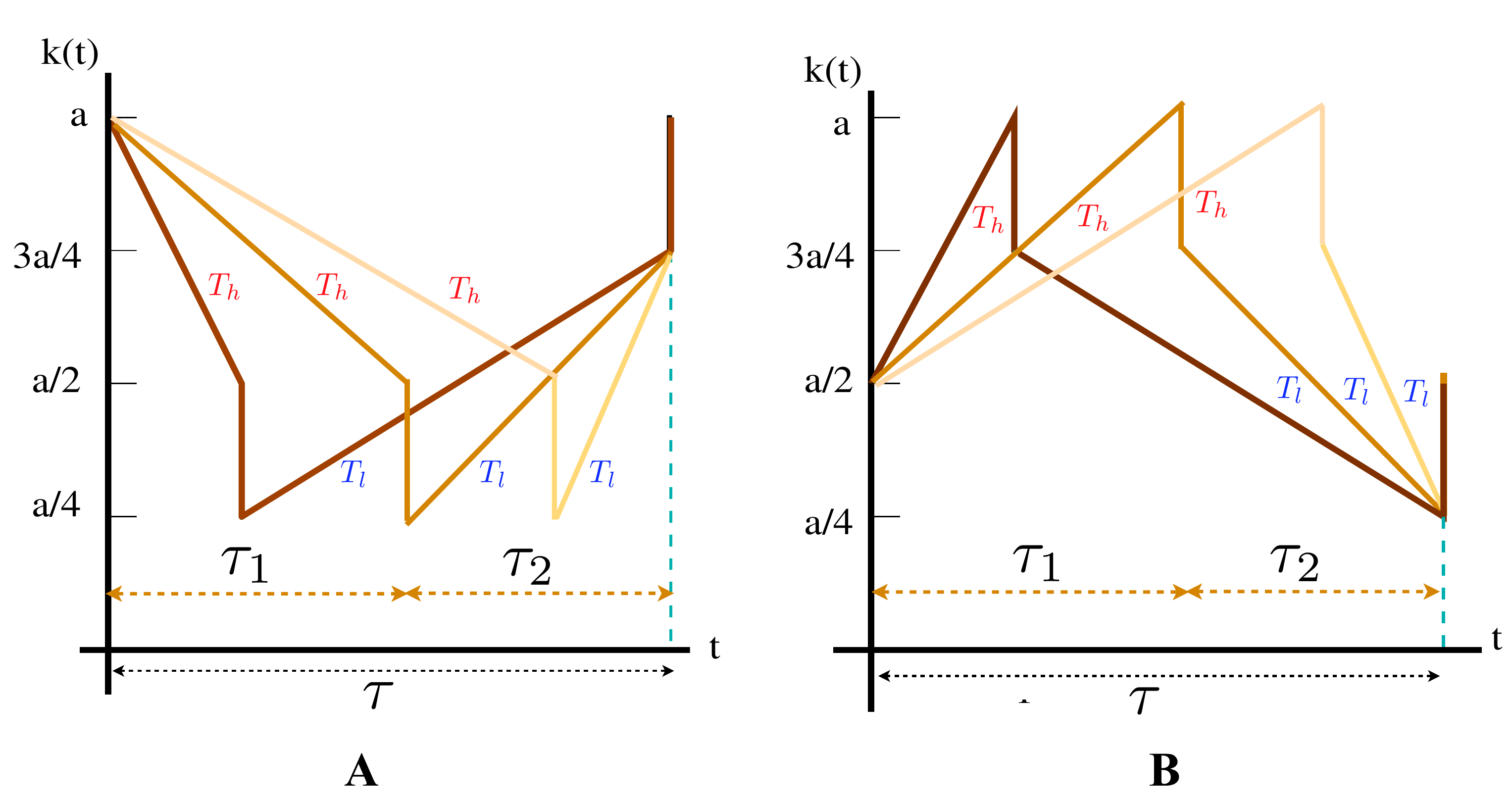}
\caption{(Color online) A. Engine protocol consisting of two isothermal steps at two different temperatures $T_h$ and $T_l$, with two instantaneous adiabatic steps connecting them. Three different protocols had been shown for three different contact time ratios($r:s$) - 1:1, 1:3 and 3:1. $\tau_1$ denotes the time during which the system undergoes an isothermal expansion in contact with the hot bath and $\tau_1$ denotes the time during which the system undergoes an isothermal compression in contact with the cold bath. B. Refrigerator protocol obtained by reversing the engine protocol. In this protocol the system undergoes isothermal compression in contact with the hot bath for a time duration of $\tau_1$ and  isothermal expansion in contact with the cold bath for a time duration of $\tau_2$.}
\label{protocol13}
\end{center}
\end{figure}

\underline{Refrigerator protocol}:  The refrigerator protocol is the reverse(in time) of the engine protocol. In the first step the system undergoes an isothermal compression in contact with the hot bath for a time duration of $\tau_1$ where the stiffness of the trap is increased from $a/2$ to $a$. Next there is adiabatic expansion process in which the stiffness of the trap is decreased instantaneously from $a$ to $3a/4$. In the third step the system go isothermal expansion process in contact with the cold bath for a duration $\tau_2$. In this step the stiffness is decreased from $3a/4$ to $a/4$. In the last step the system undergoes adiabatic compression and the stiffness is instantly changed back to $a/2$. The time dependency of stiffness is given in the following equations,
\begin{eqnarray}
 k(t)&=&\frac{1}{2}a\left(1+\frac{r+s}{r}\frac{t}{\tau}\right) \phantom x\phantom x \phantom x \phantom x\phantom x\phantom x\phantom x 0\leq t < \tau_1  \nonumber\\
 &=&3a/4 \phantom x\phantom x \phantom x \phantom x \phantom x\phantom x \phantom x \phantom x  \phantom x\phantom x\phantom x \phantom x \phantom x\phantom x \phantom x \phantom x  \phantom x \phantom x t = \tau_1\nn\\ 
 &=&a\left(\frac{3}{4}+\frac{r}{2s}-\frac{r+s}{2s}\frac{t}{\tau}\right) \phantom x\phantom x\phantom x \tau_1\leq t < \tau  \nonumber \\
 &=&a/2 \phantom x\phantom x \phantom x \phantom x \phantom x\phantom x \phantom x \phantom x \phantom x \phantom x\phantom x \phantom x \phantom x \phantom x \phantom x \phantom x  \phantom x \phantom x\phantom x t = \tau .
 \label{protocol2}
\end{eqnarray}
The protocols are shown in Fig. \ref{protocol13} .

\section{Stochastic thermodynamics}
Before we investigate further, for completeness here we describe the essentials of stochastic thermodynamics for our system. For underdamped case, using equation of motion one can write the first law along a trajectory of the particle as 
\begin{equation}
\Delta u=w-q,
\end{equation}
where $u=\frac{1}{2}mv^2+\frac{1}{2}k(t)x^2$, $w=\int \frac{\partial u}{\partial t}dt$ and $q=-\int(-\gamma v+\sqrt{\gamma T}\xi)vdt$ are change of internal energy of the particle, work done on the particle and heat exchange between the particle and the thermal bath respectively. Note that, though $w$ and $q$ depends on the points along a single trajectory of the particle but $\Delta u$ depends only on the initial and final points. 

Using the definition of work along a single trajectory one can write the work along the isothermal steps as 
\begin{eqnarray}
w_{isoth}&=&\int_0^{\tau_1} dt\frac{\partial u}{\partial t}+\int_{\tau_1}^{\tau}dt \frac{\partial u}{\partial t}  \nn\\
&=& \int_0^{\tau_1}dt\frac{1}{2}\left(\dot k x^2\right)_{T=T_h} +  \int_{\tau_1}^{\tau}dt\frac{1}{2}\left(\dot k x^2\right)_{T=T_l}. 
\label{isoth}   
\end{eqnarray}

As the adiabatic steps are instantaneous, the work along those steps is 
\begin{equation}
w_{ad}=[u(\tau_1^+)-u(\tau_1^-)]+[u(\tau^+)-u(\tau^-)]
\label{ad}
\end{equation}
and therefore the total work along a single trajectory is given by  
\begin{equation}
w=w_{isoth}+w_{ad}.
\end{equation}
The heat transfer along the isothermal paths are given by (from the first law)
\begin{equation}
q_{1}= -\int_0^{\tau_1}dt\frac{1}{2}\left(\dot k x^2\right)_{T=T_h}+[u(\tau_1^-)-u(0)] 
\label{heat1}
\end{equation} 
\begin{equation}
q_{2}= -\int_{\tau_1}^{\tau}dt\frac{1}{2}\left(\dot k x^2\right)_{T=T_l}+[u(\tau^-)-u(\tau_1^+)] .
\label{heat2}
\end{equation}  
Since the heat transfer along the adiabatic steps are zero, the total heat transfer along the total heat transfer along a cycle is $q=q_1+q_2$. Using the definition of heat and work for engine protocol one can define efficiency as 
\begin{equation}
\eta=\frac{-w}{-q_1}
\label{eta}
\end{equation}
and for refrigerator protocol COP as 
\begin{equation}
\epsilon=\frac{-q_2}{w}.
\end{equation}
$\eta$ and $\epsilon$ depend on the individual trajectory of the particle and therefore they vary stochastically for different cycles of the engine/refrigerator. Here we study their distributions in the following sections. 

Running the dynamics for large number of cycles (N), we define the average work, power and heats as 
\begin{equation}
W=\frac{1}{N}\sum_{\text{all cycles}}w; {\phantom {xx}} P=\frac{W}{\tau}; {\phantom {xx}}  {\mathcal {Q}}_1=\frac{1}{N}\sum_{\text{all cycles}} q_1; {\phantom {xx}} {\mathcal {Q}}_2=\frac{1}{N}\sum_{\text{all cycles}} q_2.
\end{equation}
Note that using ${\mathcal {Q}}_1$ and ${\mathcal {Q}}_2$, we can calculate the change of average bath entropy $\langle\Delta S_{bath}\rangle=-\frac{{\mathcal {Q}}_1}{T_h}+\frac{{\mathcal {Q}}_2}{T_l}$. In time periodic steady state and large enough $N$, 
$\langle\Delta S_{bath}\rangle=\langle\Delta S_{tot}\rangle$. Therefore from numerics we can explore non-quasistatic as well as quasistatic behaviour of thermodynamic quantities ($W$, $P$, $\mathcal {Q}_1$, $\mathcal {Q}_2$) by varying $\tau$ from small to large values. Similar analysis can be done in the overdamped limit with $u=\frac{1}{2}k(t)x^2$.

\section{Quasistatic results}

\subsection{Engine protocol}

\subsubsection{Underdamped dynamics}
We calculate the thermodynamic quantities like average work and heat exchanges for different steps of a cycle. During the isothermal processes the system instantaneously adjusts to the equilibrium state corresponding to the value of the protocol at that instant. Hence the work done along any isothermal process is the the free energy difference between the initial and the final state. In the first step, i.e., isothermal expansion the work done on the system is 
\begin{equation}
W_1=\Delta F_h=\frac{T_h}{2}\ln\left(\frac{k(t=\tau_1^-)}{k(0)}\right)=\frac{T_h}{2}\ln\left(\frac{a/2}{a}\right)=\frac{T_h}{2}\ln\left(\frac{1}{2}\right)
\label{wie}
\end{equation} 
At $t=\tau_1$, the system is in equilibrium with the bath at temperature $T_h$ with stiffness constant $a/2$. The second step being instantaneous, no heat will be dissipated and the phase space distribution given by 
\begin{equation}
P_{\tau_1}(x,v)=\frac{1}{2\pi T_h}\sqrt\frac{a}{2}\exp\left(-\frac{ax^2}{4T_h}-\frac{v^2}{2T_h}\right)
\end{equation}
remains unaltered. Correspondingly, the average work done on the particle is the change in its internal energy,
\begin{equation}
W_2=\int_{-\infty}^\infty dxdv \left(\frac{a}{4}-\frac{a}{2}\right)\frac{x^2}{2}P_{\tau_1}(x,v)=-\frac{T_h}{4}.
\label{wae}
\end{equation}
 Similarly in the isothermal compression step the work done on the particle in the quasistatic limit is 
\begin{equation}
W_3=\Delta F_l=\frac{T_l}{2}\ln\left(\frac{k(t=\tau)}{k(\tau_1)}\right)=\frac{T_l}{2}\ln\left(\frac{3a/4}{a/4}\right)=\frac{T_l}{2}\ln 3.
\label{wic}
\end{equation} 
At the end of the third step the system is in equilibrium with the cold bath with stiffness constant $3a/4$ and the corresponding distribution is 
\begin{equation}
P_{\tau}(x,v)=\frac{1}{4\pi T_l}\sqrt {3a}\exp\left(-\frac{3ax^2}{8T_l}-\frac{v^2}{2T_l}\right).
\end{equation}
In the last step, i.e., adiabatic compression step the work done on the system is 
\begin{equation}
W_4=\int_{-\infty}^\infty dxdv \left(a-\frac{3a}{4}\right)\frac{x^2}{2}P_{\tau}(x,v)=\frac{T_l}{6}.
\label{wac}
\end{equation}
 Hence, the average total work done in a full cycle of the engine protocol in the quasistatic limit is given by
\begin{equation}
W_{tot}=W_1+W_2+W_3+W_4=\frac{T_h}{2}\ln\left(\frac{1}{2}\right)-\frac{T_h}{4}+\frac{T_l}{2}\ln 3+\frac{T_l}{6}
\label{wtot_u}
\end{equation}
Heat exchanged with the hot bath in the isothermal expansion process is obtained by calculating the internal energy change and using first law of thermodynamics. At $t=0^-$, the system was in contact with the cold bath whereas at $t=0^+$ the system is connected to hot bath. Thus the system has to relax into a new equilibrium after a sudden change of temperature. This relaxation leads to a heat exchange between the system and the hot bath which accounts to change in internal energy. One can readily obtain the change in internal energy as 
\begin{equation}
\Delta U=U(\tau_1^-)-U(0^+)=T_h-\frac{7}{6}T_l.
\end{equation}  
Now, using first law, the average heat absorption from the hot bath for the first step is
\begin{equation}
-Q_1=\Delta U-W_1=T_h-\frac{7}{6}T_l-\frac{T_h}{2}\ln\left(\frac{1}{2}\right).
\label{q1_u}
\end{equation}
Similarly we can obtain the heat transferred from the cold bath to the system,
\begin{equation}
-Q_2=T_l-\frac{3}{4}T_h-\frac{T_l}{2}\ln 3.
\label{q2_u}
\end{equation}
From Eq.\ref{q1_u} and  Eq. \ref{q2_u}, $Q_1$ is negative and $Q_2$ is positive for all values of the temperature ratio $T_h/T_l$ in the quasistatic limit. But, from Eq. \ref{wtot_u}, $W$ is positive when $(T_h/T_l)< 1.2$ and negative otherwise. This implies that in the quasistatic limit, the system will act as heater II when $(T_h/T_l)< 1.2$ and as an engine when $(T_h/T_l)> 1.2$. 

The average efficiency and the average entropy production in the quasistatic limit is given by 
\begin{equation}
\eta_q=\frac{-W_{tot}}{-Q_1}=\frac{\frac{T_h}{2T_l}\ln 2+\frac{T_h}{4T_l}-\frac{1}{2}\ln 3+\frac{1}{6}}{\frac{T_h}{T_l}-\frac{7}{6}+\frac{T_h}{2T_l}\ln 2},
\end{equation}

\begin{equation}
\Delta S_q=\frac{3T_h}{4T_l}-\frac{7T_l}{6T_h}+\frac{1}{2}\ln 6,
\end{equation}

From the above expression it can be shown that $\Delta S_q$ never vanishes for any $y>1$. Hence there will be positive heat dissipation implying that the system always work in irreversible mode. Therefore the system cannot reach Carnot efficiency in the  quasistatic limit.

\subsubsection{Overdamped dynamics}
In the overdamped limit, the dynamics of the system is described by Eq. \ref{eom2} where the inertial effects are ignored. This approximation is valid when the time steps of observation is large compared to $m/\gamma$. The internal energy is given only by the potential energy term. The analytical calculations for the average thermodynamic quantities in the quasistatic limit are similar to the underdamped case. The work done in the isothermal expansion and compression are exactly the same as Eq. \ref{wie} and \ref{wic}. The work done along the adiabatic steps will be same as given by Eq. \ref{wae} and \ref{wac}, except the fact that the probability distribution will depend only on the position of the particle and not on its velocity. The total work done on the system in a whole cycle will be the same as that obtained in the underdamped case(Eq. \ref{wtot_u}). 
Using the same arguments similar to the underdamped case and keeping in mind the fact that there is only one phase space variable, namely position, the averaged internal energy change in the first step is given by 
\begin{equation}
\Delta U=U(\tau_1^-)-U(0^+)=\frac{1}{2}T_h-\frac{2}{3}T_l.
\end{equation}
Using the first law, the average heat absorption from the hot bath in the first step is 
\begin{equation}
-Q_1=\Delta U-W_1=\frac{1}{2}T_h-\frac{2}{3}T_l-\frac{T_h}{2}\ln\left(\frac{1}{2}\right)
\end{equation}
and the heat exchanged with the cold bath is given by,
\begin{equation}
Q_2=\frac{T_l}{2}\ln 3+\frac{1}{4}T_h-\frac{1}{2}T_l.
\end{equation}
From the above expressions, it is easy to show that $Q_1$ is negative and $Q_2$ is positive for all values of the temperature ratio $T_h/T_l$ in the quasistatic limit. Similar to the underdamped dynamics, the system acts in the same operational behaviour in the overdamped dynamics when the driving protocol is applied quasistatically i.e., it works as an engine when $(T_h/T_l)> 1.2$ and as heater II otherwise. 

The average efficiency and the average total entropy are given by 
\begin{equation}
\eta_q=\frac{-W_{tot}}{-Q_1}=\frac{T_h\ln 2+\frac{1}{2}T_h-T_l\ln 3-\frac{1}{3}T_l}{T_h+T_h\ln 2-\frac{4}{3}T_l},
\end{equation}
\begin{equation}
\Delta S_q=\frac{-Q_1}{T_h}+\frac{Q_2}{T_l}=\frac{1}{2}\ln 6+\frac{1}{4} \frac{T_h}{T_l}-\frac{2}{3}\frac{T_l}{T_h}.
\label{sq}
\end{equation}

\begin{figure}[H]
\begin{center}
\includegraphics[width=7cm]{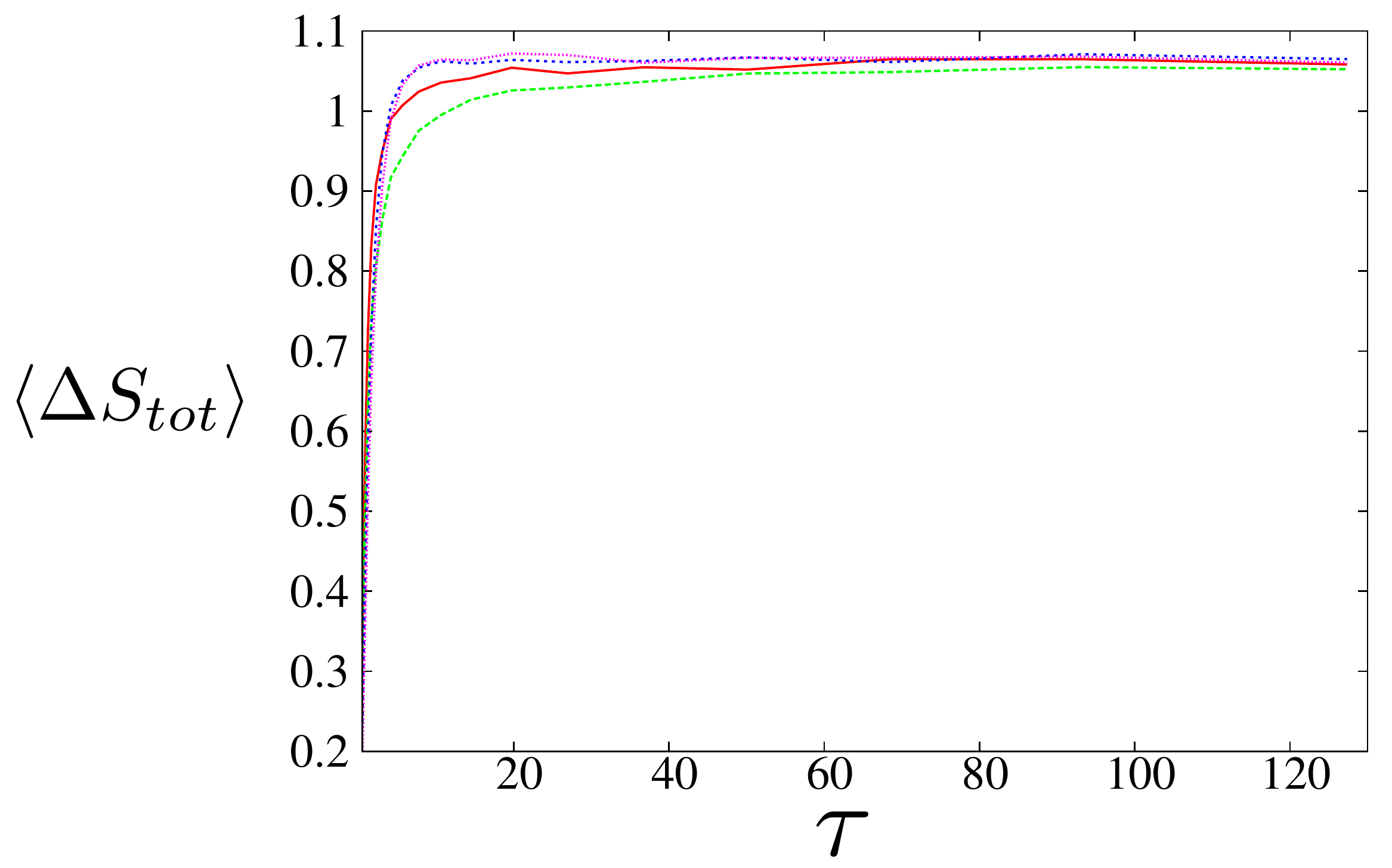}
\caption{ Average total entropy production as a function of cycle time for four different contact time ratios: red - 1:1, green - 1:3, blue - 3:1, pink - 4:1. The hot bath temperature is maintained at $T_h=0.2$. }
\label{en_dstot_ov}
\end{center}
\end{figure}

From the above expression of $\eta_q$, we conclude that for any $\frac{T_h}{T_l}>1$, $\eta_q\neq\eta_c=1-\frac{T_l}{T_h}$.  Hence the system works irreversibly. It is also in compliance with the fact that $\Delta S_q\geq 0$, evident from Eq.\ref{sq}. In fig.\ref{en_dstot_ov} we numerically calculate the total entropy with varying $\tau$ at $\frac{T_h}{T_l}=2$. Apart from the nonquasistatic behaviour of total entropy for small $\tau$, it shows that for larger $\tau$, the total entropy saturates to a nonzero positive value which is, $\Delta S_q{\Big |}_{\frac{T_h}{T_l}=2}$.   

\begin{figure}[H]
\begin{center}
 \includegraphics[width=7cm]{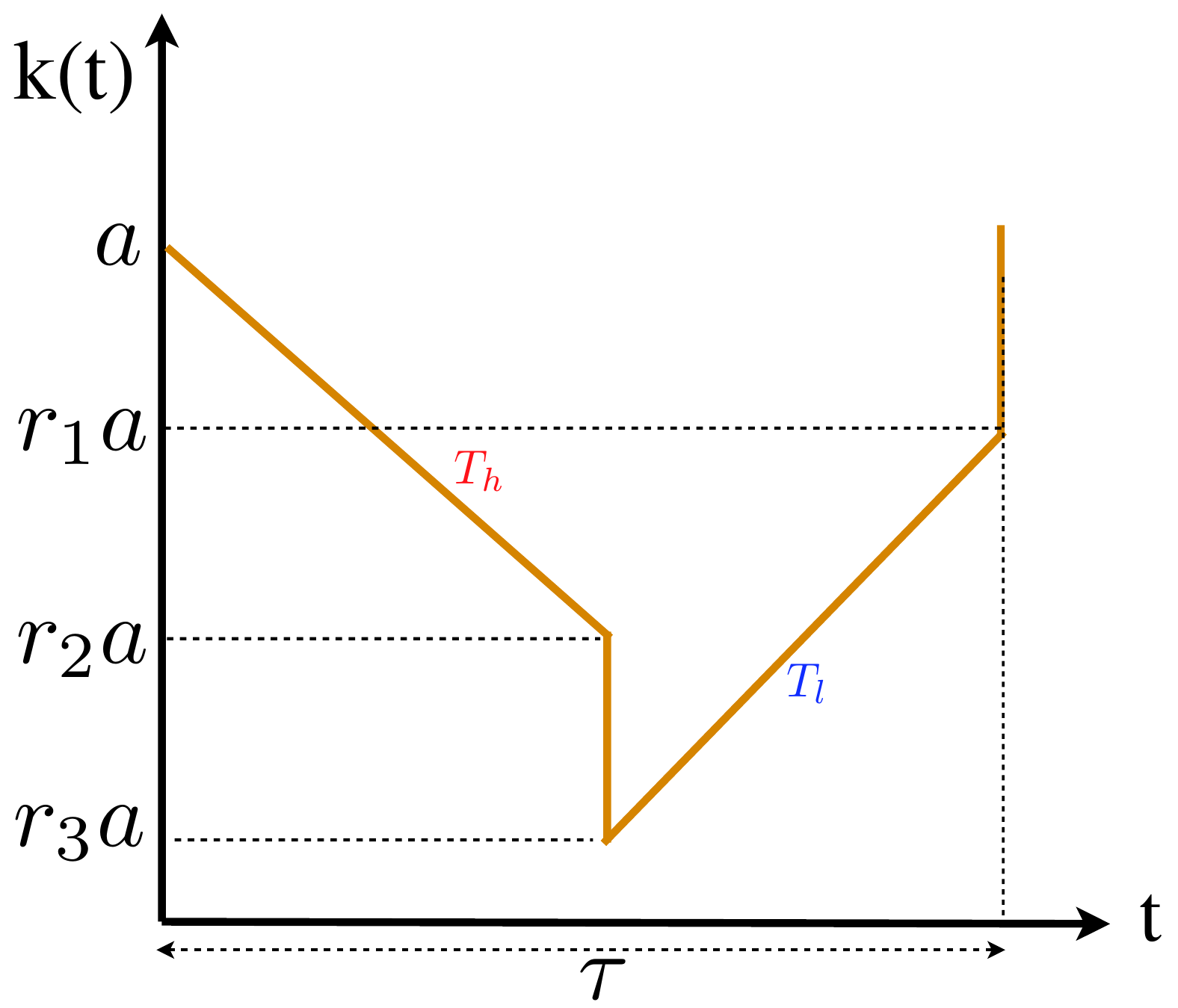}
\caption{(Color online)Carnot type engine protocol with arbitrary jump heights. The system first undergoes an isothermal expansion in presence of hot bath where the stiffness of the trap is changed from $a$ to $r_2a$. In the second step the system is subjected to adiabatic expansion by changing the stiffness instantaneously from $r_2a$ to $r_3a$. Next the stiffness is changed isothermally changed in presence of cold bath from $r_3a$ to $r_1a$. In the final step, the stiffness of the is suddenly changed from $r_1a$ to $a$. One should note that $r_3<r_2<r_1<1$. In order to get a reversible mode in the quasistatic limit one needs to main a certain temperature ratio $\left(\frac{T_h}{T_l}\right)$ and $r_1, r_2  \& r_3$ should maintain a definite ratio(derived in the text). As this is a quasistatic result one should not consider about the time-asymmetry of the protocol. That is why the protocol shown in the figure is time-symmetric but have different jump heights.}
\label{protocol12}
\end{center}
\end{figure}
As opposed to the protocols in \cite{rana14, rana16}, our protocol does not exhibit Carnot efficiency for any values of $T_h$ and $T_l$ in the overdamped quasistatic limit. This is due to the equal adiabatic jump heights in our protocol.  One can change the jump heights and obtain the Carnot value given the fact that the heights and the temperature of the two baths maintain a certain relation. To demonstrate this, let's consider a general protocol as shown in Fig. \ref{protocol12} and calculate the efficiency in the quasistatic regime. 

The work done in the first isothermal step when the system is attached to hot bath at temperature $T_h$ is given by 

\begin{equation}
W_1=\frac{T_h}{2}\ln\left[\frac{k(t=\frac{\tau}{2}^-)}{k(0)}\right]=\frac{T_h}{2}\ln\left[\frac{r_2a}{a}\right]=\frac{1}{2}T_h\ln(r_2)
\end{equation}

At the end of the first step the system is in thermal equilibrium with the hot bath with stiffness $r_2a$. The average work done in the second step is given by average internal energy change,
\begin{equation}
W_2=\int_{-\infty}^\infty dx \frac{1}{2}(r_3a-r_2a)x^2\sqrt{\frac{r_2a}{2\pi T_h}}\exp\left(-\frac{r_2a}{2T_h}x^2\right)=\frac{1}{2r_2}(r_3-r_2)T_h
\end{equation}

Work done in the third step i.e., when the system is undergoing an isothermal process in contact with the cold bath is 
\begin{equation}
W_3=\frac{T_l}{2}\ln\left[\frac{k(t=\tau^-)}{k(\frac{\tau}{2})}\right]=\frac{T_l}{2}\ln\left[\frac{r_1a}{r_3a}\right]=\frac{1}{2}T_l\ln\left(\frac{r_1}{r_3}\right)
\end{equation}

The last step being adiabatic compression step, the stiffness of the harmonic trap is changed instantaneously from $r_1a$ to $a$ while the system is still in equilibrium with the cold bath. As a result the heat dissipation vanishes and the work done is readily given by the internal energy change 

\begin{equation}
W_4=\int_{-\infty}^\infty dx \frac{1}{2}(a-r_1a)x^2\sqrt{\frac{r_1a}{2\pi T_l}}\exp\left(-\frac{r_1a}{2T_l}x^2\right)=\frac{1}{2r_1}(1-r_1)T_l.
\end{equation} 
Hence the average total work in the quasistatic limit is given by 
\begin{eqnarray}
W_{tot}&=&W_1+W_2+W_3+W_4\nonumber\\
&=&\frac{1}{2}T_h\ln(r_2)+\frac{1}{2r_2}(r_3-r_2)T_h+\frac{1}{2}T_l\ln\left(\frac{r_1}{r_3}\right)+\frac{1}{2r_1}(1-r_1)T_l
\end{eqnarray}
Using the similar arguments as before one can calculate the internal energy change in the isothermal expansion as $\Delta U=\frac{1}{2}\left(T_h-\frac{1}{r_1}T_l\right)$. The average heat absorbed from the hot bath in this step is given by 
\begin{equation}
-Q_1=\Delta U-W_1=\frac{1}{2}\left(T_h-\frac{1}{r_1}T_l\right)-\frac{1}{2}T_h\ln(r_2).
\end{equation}

The average efficiency and the average entropy production  in the quasistatic regime is given by 
\begin{equation}
\eta_q=\frac{-W_{tot}}{-Q_1}=\frac{\frac{T_h}{T_l}\ln(r_2)+\frac{T_h}{T_lr_2}(r_3-r_2)+\ln\left(\frac{r_1}{r_2}\right)+\frac{1}{r_1}(1-r_1)}{\frac{T_h}{T_l}\ln(r_2)-\frac{T_h}{T_l}+\frac{1}{r_1}},
\end{equation}

\begin{equation}
\Delta S_q=\frac{-Q_1}{T_h}+\frac{Q_2}{T_l}=1-\frac{T_h}{2T_l}-\frac{1T_l}{2r_1T_h}-\frac{1}{2r_2}(r_3-r_2)\frac{T_h}{T_l}-\frac{1}{2}\ln\left(\frac{r_1r_2}{r_3}\right).
\end{equation}

From the above expression, it can be easily shown that $\eta_q$ equals Carnot efficiency $\eta_c\left(=1-\frac{T_l}{T_h}\right)$ when  both the conditions namely $\frac{T_h}{T_l}=\frac{r_2}{r_3}$ and $r_1=\frac{r_3}{r_2}$ are satisfied.  
Under these conditions the entropy production vanishes and hence the system works in reversible mode \cite{sek00}.

\subsection{Refrigerator protocol}

\subsubsection{Underdamped dynamics}
Here we calculate the average thermodynamical quantities in quasistatic limit under the application of refrigerator protocol. In the first step, i.e., isothermal compression step the work done on the particle is the free energy difference as given by,
\begin{equation}
W_1'=\Delta F_h=\frac{T_h}{2}\ln\left(\frac{k(\tau_1^-)}{k(0)}\right)=\frac{T_h}{2}\ln\left(\frac{a}{a/2}\right)=\frac{T_h}{2}\ln 2
\end{equation}
At $t=\tau_1^-$, the system is in equilibrium with the hot bath and the corresponding probability distribution is given by
\begin{equation}
P_{\tau_1}(x,v)=N_1'\exp\left[-\left(\frac{ax^2}{2T_h}+\frac{v^2}{2T_h}\right)\right],
\end{equation}
where $N_1'=\frac{\sqrt a}{2\pi T_h}$. The second is an adiabatic expansion process. This step is instantaneous and hence no heat is dissipated to the bath. So, the work done on the particle is the instantaneous change in its internal energy, given by
\begin{equation}
W_2'=\int_{-\infty}^{\infty} \frac{1}{2}\left(\frac{3a}{4}-a \right)x^2P_{\tau_1}(x,v) dxdv=-\frac{1}{8}T_h
\end{equation}
Similar to the first step, the work done on the particle in the isothermal expansion(third step) is given by,
\begin{equation}
W_3'=\frac{T_l}{2}\ln\left(\frac{k(\tau^-)}{k(\tau_1^+)}\right)=\frac{T_l}{2}\ln\left(\frac{a/4}{3a/4}\right)=\frac{T_l}{2}\ln\left(\frac{1}{3}\right).
\end{equation}
At the end of the third step, the system is in equilibrium with the cold bath and probability distribution of the state of the particle is,
\begin{equation}
P_{\tau}(x,v)=N_2'\exp\left[-\left(\frac{ax^2}{8T_l}+\frac{v^2}{2T_l}\right)\right],
\end{equation}
where $N_2'=\frac{\sqrt a}{4\pi T_l}$. In the last step, the average work done on the particle is given by
\begin{equation}
W_4'=\int_{-\infty}^{\infty} \frac{1}{2}\left(\frac{a}{2}-\frac{a}{4}\right)x^2P_{\tau}(x,v) dxdv=\frac{1}{2}T_l.
\end{equation}
The total average work is given  by 
\begin{equation}
W_{tot}^{ref}=W_1'+W_2'+W_3'+W_4'=\frac{T_h}{2}\ln 2-\frac{1}{8}T_h+\frac{T_l}{2}\ln\left(\frac{1}{3}\right)+\frac{1}{2}T_l.
\label{wtot_ru}
\end{equation}
To obtain heat absorption from the cold bath ($Q_2^{ref}$)  first we have to calculate internal energy change along the third step. The average  internal energy at $t=\tau_1^+ $  is $U\left(\frac{\tau}{2}^+\right)=
\int_{-\infty}^{\infty}\left(\frac{3a x^2}{8}+\frac{v^2}{2}\right)P_{\tau_1}(x,v)dxdv=\frac{7}{8}T_h$. Since the system is in equilibrium with the cold bath at $t=\tau^-$, the average internal energy will be $T_l$. This leads to the change in internal energy in the third step, $\left(T_l-\frac{7}{8}T_h\right) $. Using first law, we obtain average heat dissipated to the cold bath, 
\begin{equation}
Q_2^{ref}=\frac{T_l}{2}\ln\left(\frac{1}{3}\right)-T_l+\frac{7}{8}T_h.
\label{q2_ru}
\end{equation}
Similarly one can obtain the heat transferred to the hot bath,
\begin{equation}
Q_1^{ref}=\frac{T_h}{2}\ln 2-T_h+\frac{3}{2}T_l.
\label{q1_ru}
\end{equation} 
It can be easily shown from Eq. \ref{wtot_ru}, $W_{tot}^{ref}$ is positive for all values of $T_h/T_l$ i.e., work is always done on the system in the quasistatic limit. But Eq. \ref{q2_ru} and Eq. \ref{q1_ru} gives us three regimes where the system acts in three different operational mode depending on $(T_h/T_l)$. When $(T_h/T_l)<1.77$, $Q_2^{ref}$ is negative and $Q_1^{ref}$ is positive and hence the system act as a refrigerator in the quasistatic limit. When $1.77<(T_h/T_l)<2.29$, both $Q_2^{ref}$ and $Q_1^{ref}$ are positive. Under this condition the system works as heater I. When $2.29<(T_h/T_l)$, $Q_2^{ref}$ is positive and $Q_1^{ref}$ is negative and the system behaves as heater II.

\subsubsection{Overdamped dynamics}
Quasistatic calculations similar to the underdamped case can also be done for overdamped limit with only potential energy contributing to the total internal energy. The total average work done in a cycle in the overdamped limit is same as that obtained for the underdamped case. At $t=\tau_1^+$, the average internal energy is given by $U(\tau_1^+)=\frac{3}{8}T_h$. At $t=\tau^-$, the system is in equilibrium with cold bath and the corresponding average internal energy is $U(\tau^-)=\frac{1}{2}T_l$. Hence the change in internal energy in the isothermal expansion process is $\left(\frac{1}{2}T_l-\frac{3}{8}T_h\right)$. Hence, the average heat that is transferred to the cold bath  in the quasistatic limit is given by
\begin{equation}
Q_2^{ref}=\frac{T_l}{2}\ln\left(\frac{1}{3}\right)-\frac{1}{2}T_l+\frac{3}{8}T_h.
\label{q2_ro}
\end{equation}
The heat exchanged between the hot bath and the system is given by 
\begin{equation}
Q_1^{ref}=\frac{T_h}{2}\ln 2-\frac{1}{2}T_h+T_l.
\label{q1_ro}
\end{equation}
Similar to the underdamped case the work done on the system in overdamped  dynamics is always positive. Eq. \ref{q2_ro} and Eq. \ref{q1_ro} gives us three domains, depending on the ratio of the hot and the cold bath temperatures, where the system works in three different operational modes. When $(T_h/T_l)<2.79$, $Q_1^{ref}$ is positive and $Q_2^{ref}$ is negative i.e., heat is carried from cold bath to the hot bath using the work done on the system. Under this condition the system works as a refrigerator in the quasistatic limit. When $2.79<(T_h/T_l)<6.51$, both $Q_1^{ref}$ and $Q_2^{ref}$ are positive i.e., work done on the system is being used to heat up both the baths and the system acts as a heater I. When $6.51<(T_h/T_l)$, $Q_1^{ref}$ is negative and $Q_2^{ref}$ is positive. In this temperature regime the system performs as heater II. 

\section{Numerical results}
In this section we explore the non-quasistatic regime. We evolve the system using discretised Langevin dynamics with time step $dt=0.001$ in the underdamped as well as overdamped limit [Eq. \ref{eom1} and Eq. \ref{eom2}]. The system is driven by time periodic protocols [Eq. \ref{protocol1} and Eq. \ref{protocol2}]. We follow Heun's method \cite{heun}. We have set $\gamma=1$ and $m=1$. All the physical quantities are in dimensionless form. Throughout the paper we have fixed $a=5$ and $T_l=0.1$. We have considered four different values of the ratio $r:s$ namely $1:1$, $1:3$, $3:1$ and $4:1$ and compared the results between them. We made sure that, after the initial transient regime($\sim 10^3$ cycle time), the system settles to a TPSS i.e., $P_{ss}(x,v,t+\tau)=P_{ss}(x,v,t)$. 

\subsection{Engine Protocol}

\subsubsection{Underdamped dynamics}
\begin{figure}[H]
\begin{center}
\includegraphics[width=12cm]{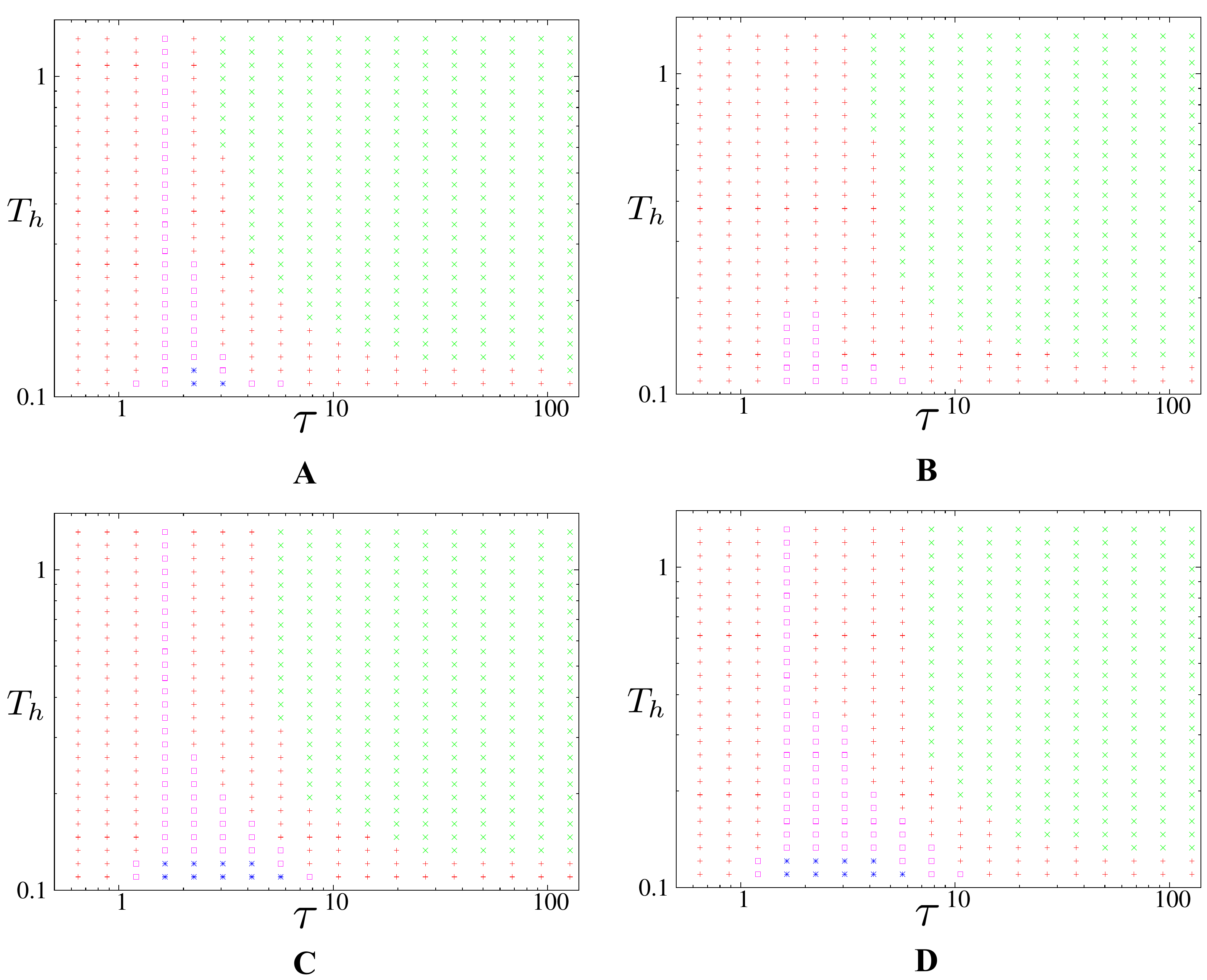}
\caption{ Different modes of operation of our system following underdamped Langevin dynamics with Carnot engine protocol:  Open boxes(pink): heater-I, asterisk(blue): refrigerator, cross(green): engine, plus(red): heater-II. The ratio ($r:s$) of the contact time with hot bath during the isothermal expansion to the contact time with the cold bath  during the isothermal compression are:  A - 1:1, B - 1:3, C - 3:1 and D - 4:1.}
\label{eu}
\end{center}
\end{figure}

{\underline{Phase diagram}}: For each $(\tau,T_h)$ pair, we calculate $W$, ${\mathcal{Q}}_1$, and ${\mathcal{Q}}_2$ and thereby obtained the phase diagram [Fig. \ref{eu}] for the operational modes of the system for four different asymmetric protocols.  From these phase diagrams it is clear that the area of different regions changes depending on the asymmetry of the protocol. 
In Fig. \ref{eu} A the phase diagram is shown for $r:s$ equals to $1:1$ i.e., the driving protocol is symmetric. Under the effect of this protocol the system acts as engine or heater II in the quasistatic limit depending on the hot bath temperature. In non-quasistatic  regime the system can act as heater I or even as a refrigerator depending on $\tau$ and $T_h$. One point to be noted, here, that the system behaves in the same way in the quasistatic limit irrespective of the asymmetry in the driving protocol and hence the phase diagram is similar in all the plots in Fig.\ref{eu} in the quasistatic limit. This quasistatic limit is consistent with our analytical results as mentioned earlier. In the non-quasistatic regime the plots are different. In Fig. \ref{eu} B, the phase diagram for  the asymmetry ratio $r:s$ equals to 1:3 i.e., the isothermal compression takes place for larger time than the isothermal expansion. In this case the refrigerator region vanishes and the region of heater I diminishes. On the other hand when the isothermal expansion takes place for a larger time than the isothermal compression the refrigerator and the heater I region increases as evident from Fig. \ref{eu} C and D, where the asymmetry ratio is 3:1 and 4:1 respectively. 

\begin{figure}[H]
\begin{center}
\includegraphics[width=7cm]{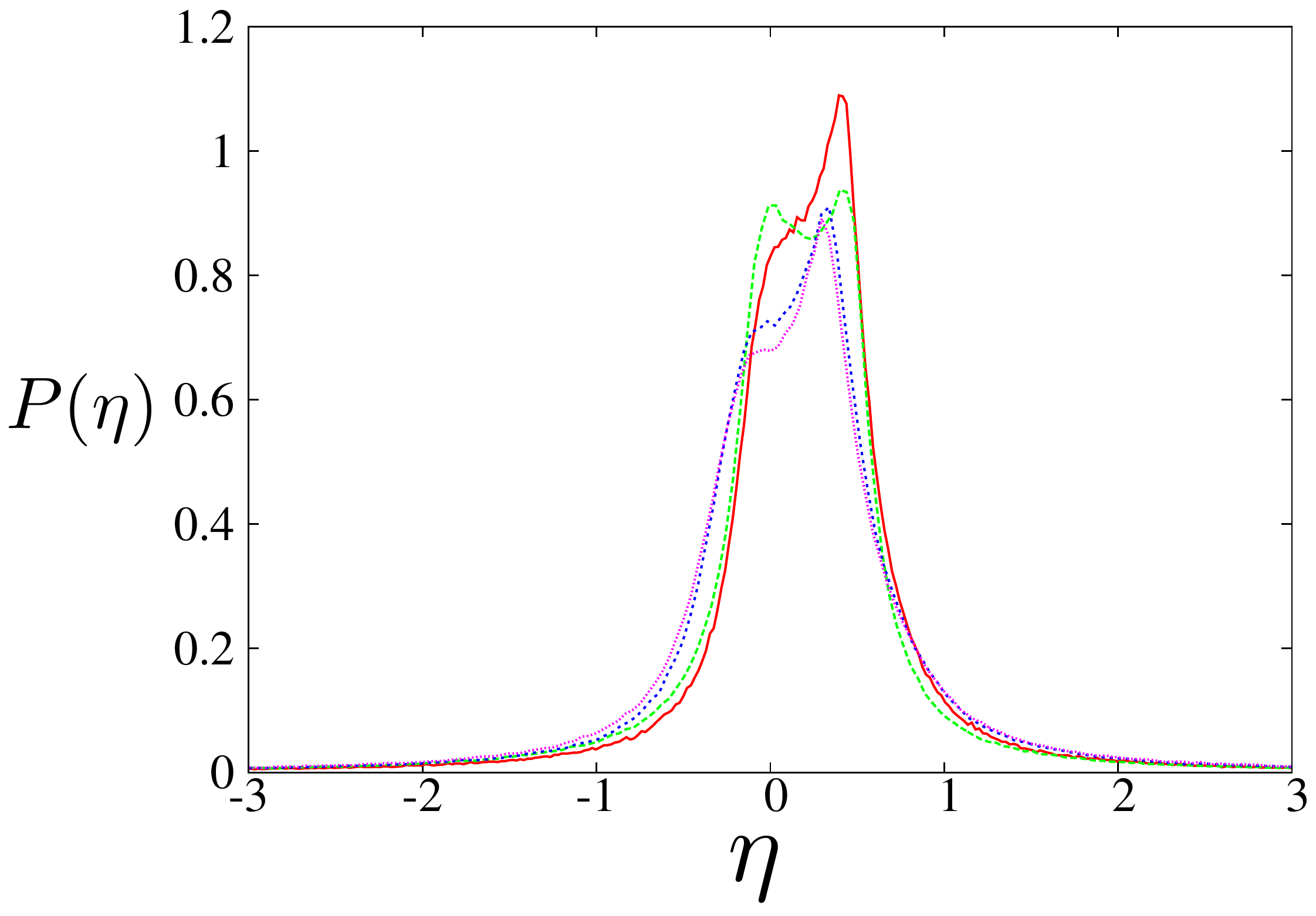}
\caption{ Efficiency distribution for four different contact time ratios: red - 1:1, green - 1:3, blue - 3:1, pink - 4:1 at $\tau=10$ and $T_h=0.5$.}
\label{eta_un}
\end{center}
\end{figure}

{\underline{Efficiency distribution}}: 
The effect of time-asymmetry in the protocols is evident in efficiency statistics. In Fig.\ref{eta_un} we have plotted the distribution of efficiency in the non-quasistatic regime with cycle time $\tau=10$ and hot bath temperature $T_h=0.5$.  Efficiency, being a ratio of two stochastic quantities namely work and heat, can take any value ranging from $-\infty$ to $+\infty$. More importantly we  see power law ($\sim \eta^{-\alpha}$) tails in the distribution both in positive and negative side with exponent $\alpha\simeq 2$. A point is to be noted from this plot is that the mean value is dominated by the standard deviation of the distribution thereby making efficiency a non-self averaging quantity. Apart from the aforesaid general properties of all the plots in Fig.\ref{eta_un}, one can also see the influence of the time-asymmetry . When the system undergoes compression for large part of the cycle time, i.e., in $1:3$ case, the distribution exhibits a double peak behaviour. This behaviour gradually changes when the time for compression is decreased and the expansion time is increased. The double peak changes into a single peak with a shoulder on the negative side of the $\eta$- axis.

\begin{figure}[H]
\begin{center}
\includegraphics[width=6cm]{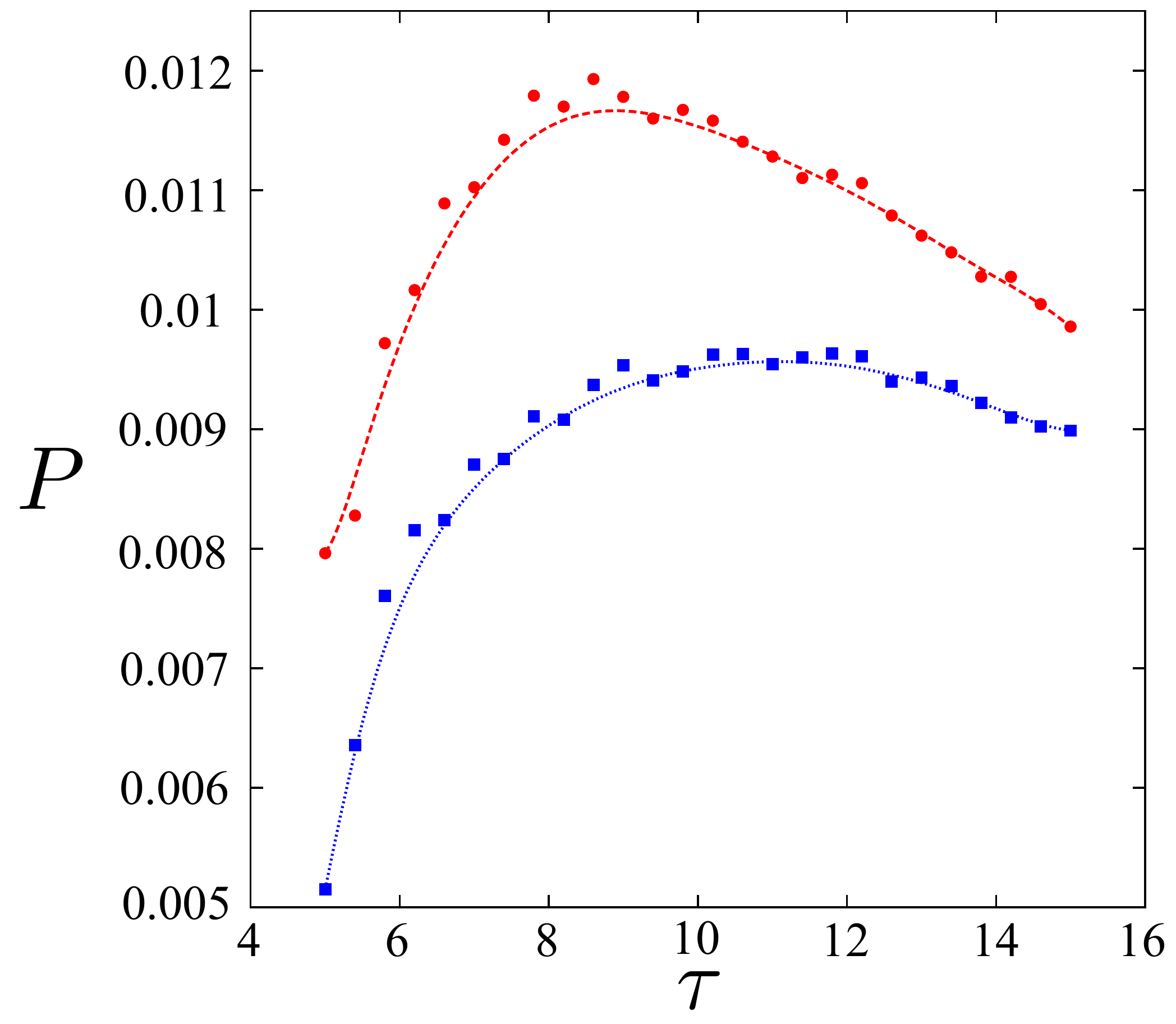}
\caption{ Power with cycle time for two different contact time ratios: red - 1:1 and blue - 3:1 at $\tau=10$ and $T_h=0.5$. The corresponding dashed lines are guide to eye.}
\label{power}
\end{center}
\end{figure}

{\underline{Power}}:  
The asymmetry in the protocol also affects the power of the engine. This fact can be seen in Fig. \ref{power}, where average power($P$) versus $\tau$ are plotted for two different asymmetry ratios($r:s$) - 1:1 and 1:3. The major difference between the asymmetric and symmetric drives is the maximum power of the symmetric protocol is larger than that of the asymmetric one and shifts towards larger value of cycle time. In the limit of $\tau\rightarrow\infty$, the power tends to zero (not shown in the figure). Efficiency at maximum power for symmetric protocol is $\eta_{max. P}=0.212$. It is much below than the Curzon-Ahlborn(C-A) bound  $\eta^{CA}=1-\sqrt(\frac{T_l}{T_h})=0.55$. Similar results hold for other asymmetric parameters. Thus we find no correlation between $\eta_{max. P}$ and C-A bound.

\subsubsection{Overdamped dynamics}
\begin{figure}[H]
\begin{center}
\includegraphics[width=12cm]{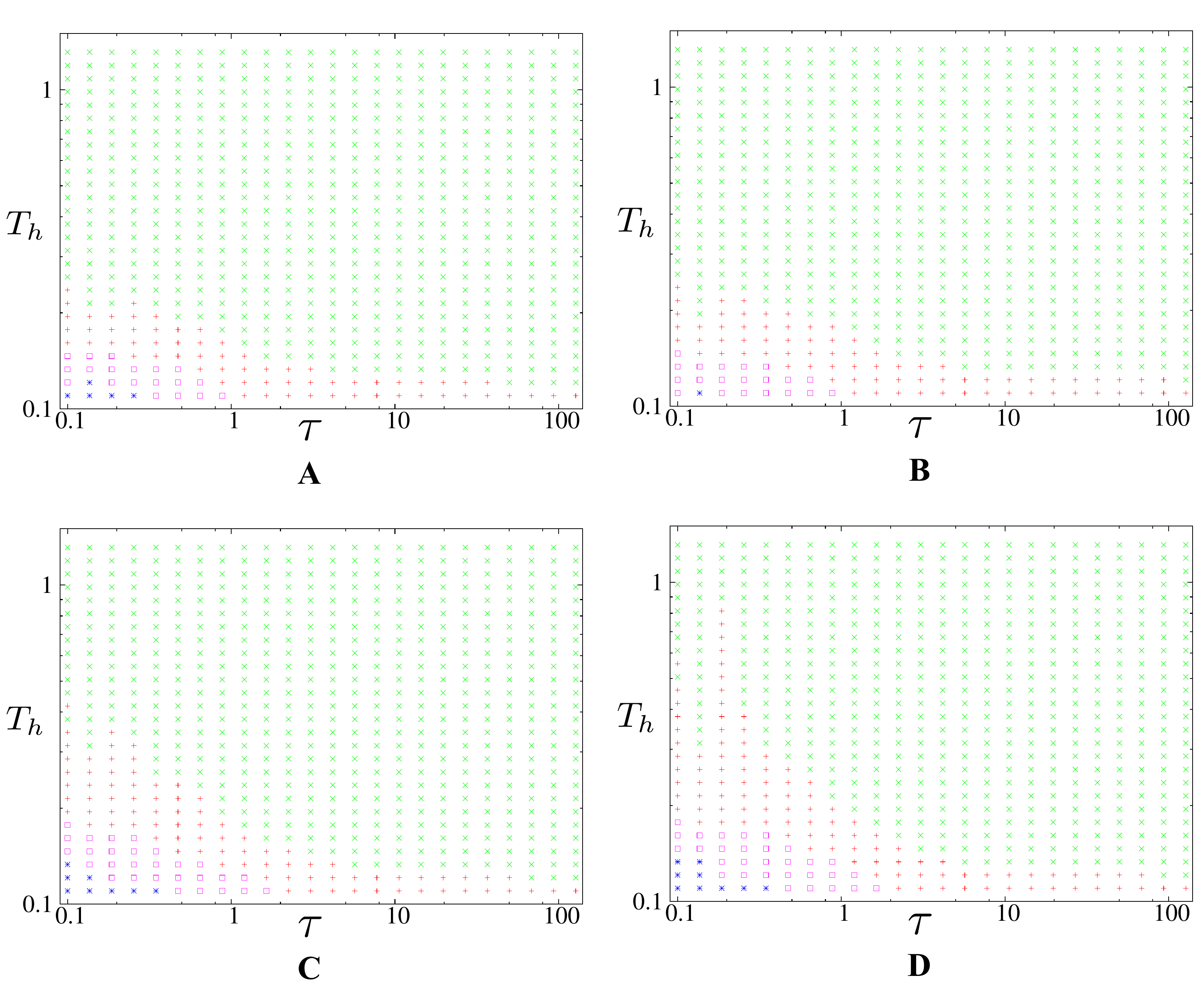}
\caption{ Different modes of operation of our system following overdamped Langevin dynamics with Carnot engine protocol:  Open boxes(pink): heater-I, asterisk(blue): refrigerator, cross(green): engine, plus(red): heater-II. The ratio ($r:s$) of the contact time with hot bath during the isothermal expansion to the contact time with the cold bath  during the isothermal compression are:  A - 1:1, B - 1:3, C - 3:1 and D - 4:1.}
\label{eo}
\end{center}
\end{figure}

{\underline{Phase diagram}}: We have scanned the parameter space ($\tau-T_h$), keeping $T_l$ fixed at 0.1, to obtain the phase diagram in overdamped regime (Fig. \ref{eo}). It clearly depicts four different modes of operation of the system. 
 In contrast to the underdamped case, no critical cycle time is required for the operation in engine mode. Therefore, total phase space area of the endine mode has increased in this limit. The phase boundaries in quasistatic limit are consistent with our analytical results. Asymmetry in driving protocol has affected the non-quasistatic part of the phase diagram. When the system is driven by a protocol with $\tau_1<\tau_2$ i.e., the time for isothermal expansion is smaller than that of isothermal compression, the refrigerator region vanishes as shown in Fig. \ref{eo} B. Fig. \ref{eo} C and Fig. \ref{eo} D displays the phase diagram when $\tau_1>\tau_2$. Thus phase diagram is sensitive to asymmetry ratio.
\begin{figure}[H]
\begin{center}
\includegraphics[width=7cm]{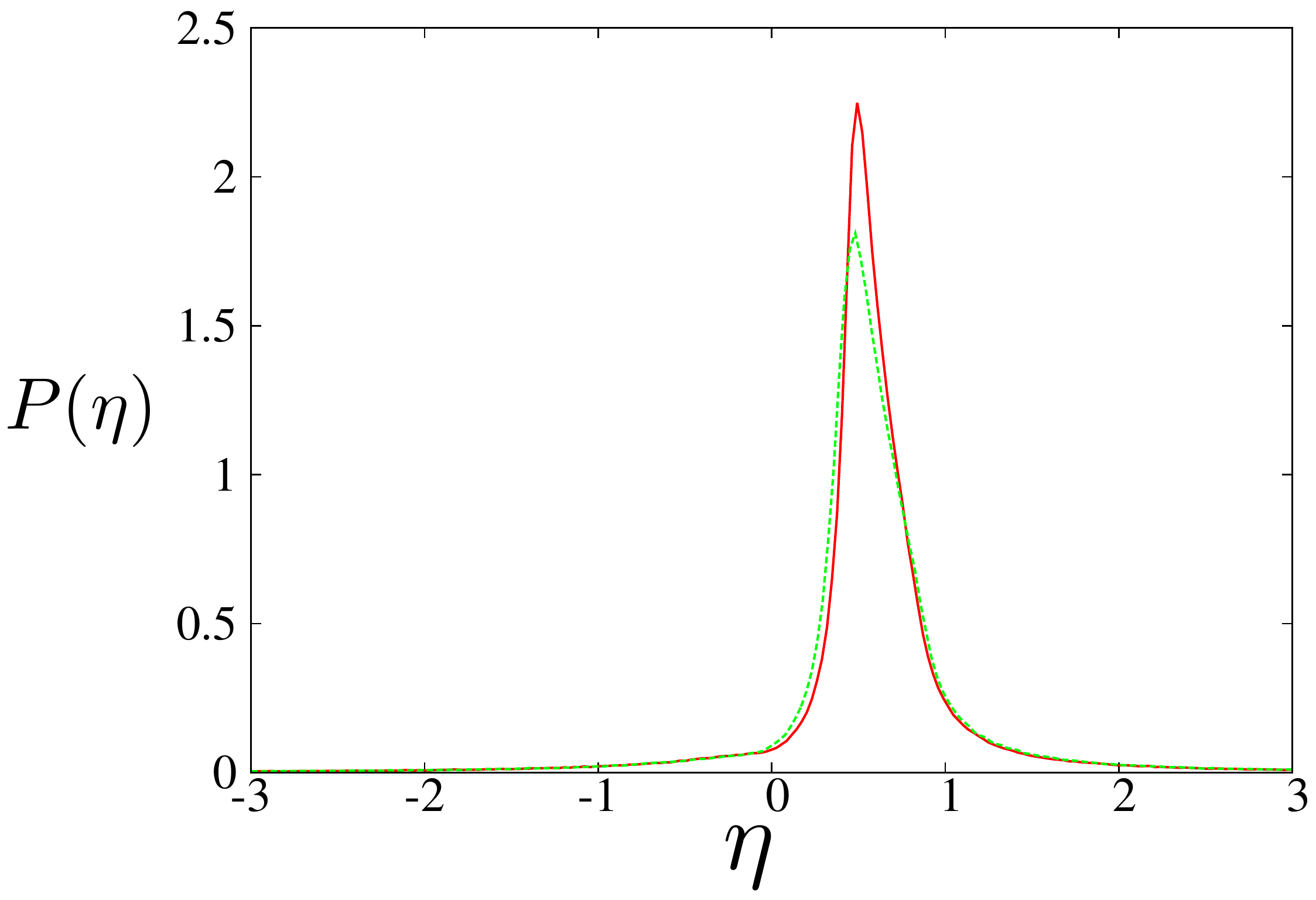}
\caption{ Efficiency distribution for two different contact time ratios: red - 1:1 $\&$ green - 3:1 at $\tau=10$ and $T_h=0.5$.}
\label{eta_ov}
\end{center}
\end{figure} 
{\underline{Efficiency distribution}}: In Fig.\ref{eta_ov}, efficiency distribution in the overdamped limit is plotted for two asymmetry ratio at $\tau=10$ and $T_h=0.5$. It is evident from the plot, that due to stochastic behaviour of the system, there exists  trajectories for which $\eta<0$ and even greater than Carnot bound($\eta=0.8$ for this case). At large values of $\eta$, the probability distribution $P(\eta)$ decays as power law ($\sim \eta^{-\alpha}$) with $\alpha\sim 2$. The effect of time asymmetry of the protocol is not significant in high damping limit. 

\subsection{Refrigerator Protocol}

\subsubsection{Underdamped dynamics}
\begin{figure}[H]
\begin{center}
\includegraphics[width=12cm]{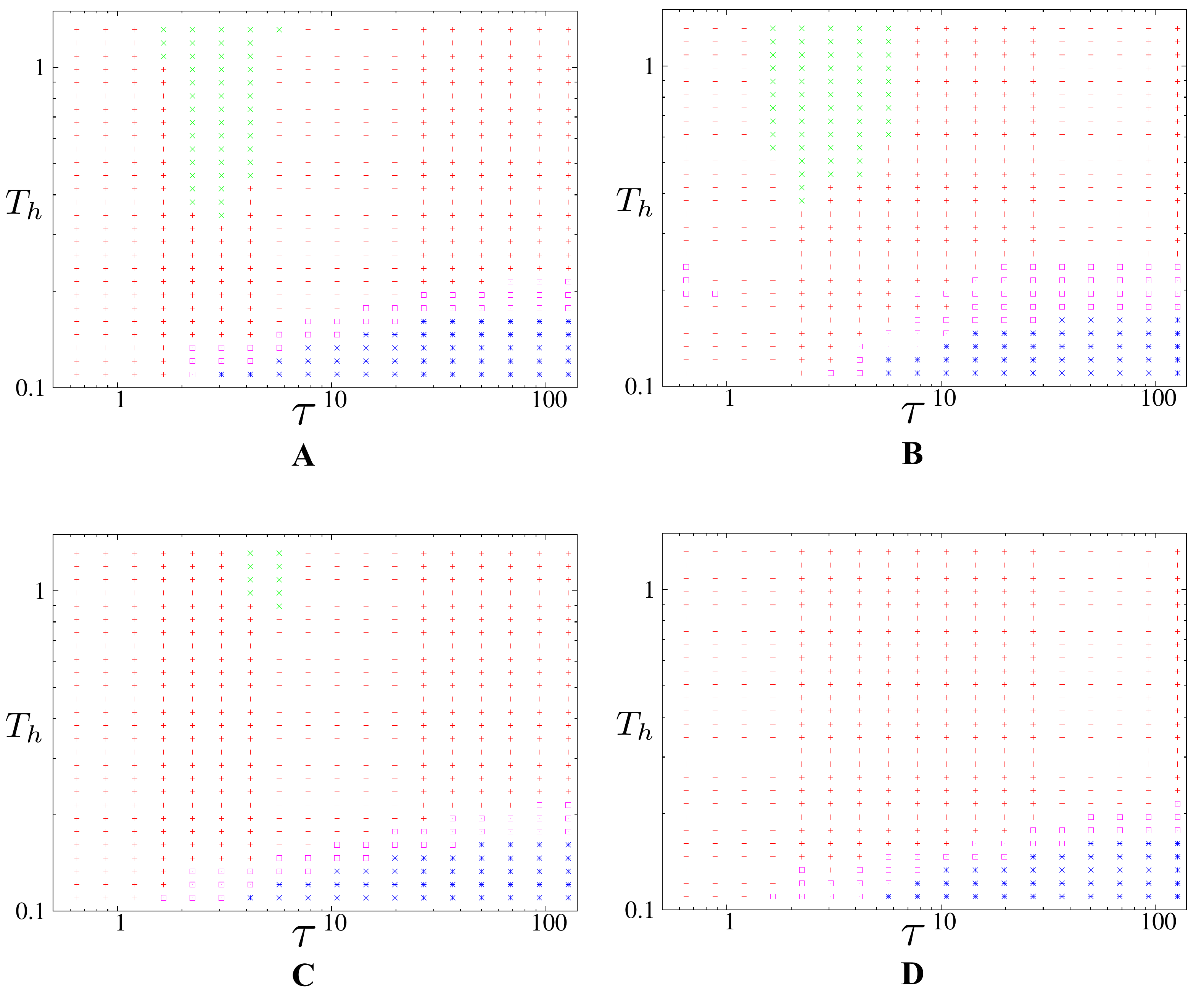}
\caption{ Different modes of operation of our system following underdamped Langevin dynamics with Carnot refrigerator protocol:  Open boxes(pink): heater-I, asterisk(blue): refrigerator, cross(green): engine, plus(red): heater-II. The ratio ($r:s$) of the contact time with hot bath during the isothermal compression to the contact time with the cold bath  during the isothermal expansion are:  A - 1:1, B - 1:3, C - 3:1 and D - 4:1.}
\label{ru}
\end{center}
\end{figure}

{\underline{Phase diagram}}: In Fig. \ref{ru}, we have shown the phase diagram of the operation of our system. The four plots in this figure correspond to protocols with different asymmetric ratio $r:s$. Even though the quasistatic behaviour is same for all cases, the non-quasistatic behaviour is highly affected due to asymmetry. In Fig. \ref{ru} A, we plotted the phase diagram when there is no asymmetry in the protocol i.e., $r:s=1:1$. It shows that one needs a critical cycle time above which the system can act as a refrigerator otherwise it performs as a heater of type II.  Another interesting point is that apart from acting as refrigerator, heater I and heater II, our system can  even work as engine in some parameter space ($T_h\gtrsim 0.3, 1.5\lesssim \tau\lesssim 6$). 
This engine region in the phase diagram can be tampered by involving asymmetry in the protocol. In Fig. \ref{ru} B, we have used a protocol with asymmetry ratio $1:3$ and obtained the phase diagram.  It is clear from the plot that the engine region  has increased when the isothermal expansion takes place for longer time than isothermal compression. In Fig. \ref{ru} C and D, we have plotted the phase diagram for asymmetry ratio $3:1$ and $4:1$ respectively. These two plots clearly depicts the fact that when the time of isothermal compression is increased more than the isothermal expansion, the engine region gradually vanishes. This is quite evident from the following fact. If the compression step takes place for larger time than expansion step, more work will be done on the system. Work extraction is small due to short duration of expansion step. So, in the whole cycle net work extraction in not possible. Under this circumstances, the system will work as either of the two types of heater.

\begin{figure}[H]
\begin{center}
\includegraphics[width=7cm]{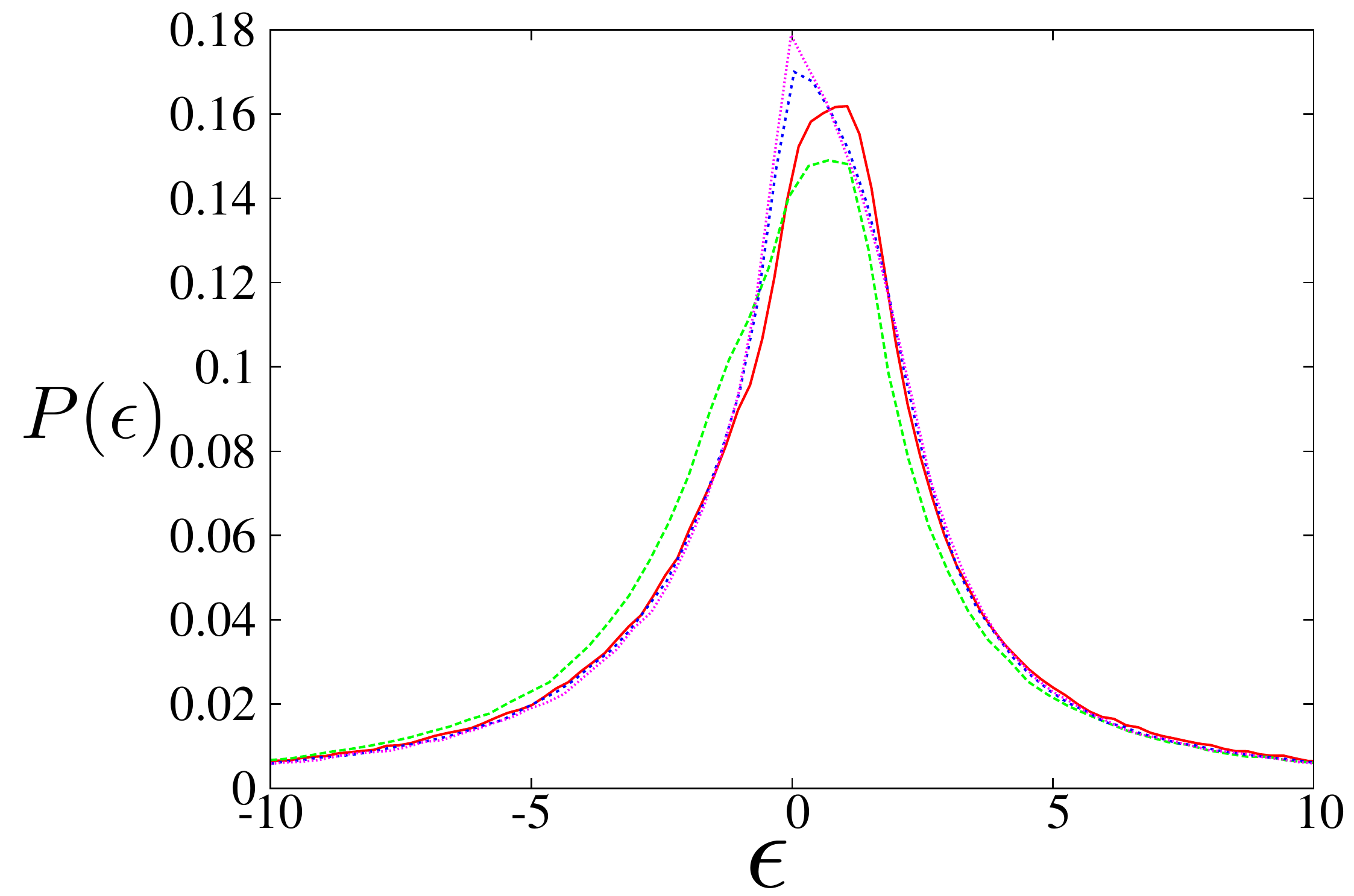}
\caption{ Distribution of COP for four different contact time ratios: red - 1:1, green - 1:3, blue - 3:1, pink - 4:1 at $\tau=10$ and $T_h=0.12$.}
\label{coeff_un}
\end{center}
\end{figure}

{\underline{COP distribution}}: $P(\epsilon)$, the distribution of stochastic COP, is shown in  Fig. \ref{coeff_un} for $\tau=10$ for four different contact time ratios. COP can take any value between $-\infty$ to $+\infty$. There are considerable number of trajectories where COP can occur beyond Carnot bound. It can even be negative for significant number of realisations. Moreover, we have noticed that the tail of $P(\epsilon)$ contains largely deviated values and has power law  ($\sim \epsilon^{-\alpha}$) decay for several decades. The value of $\alpha$ is $2\pm 0.1$. Like efficiency, the mean value is dominated by the standard deviation of the distribution thereby making efficiency a non-self averaging quantity and one need the full distribution of $\epsilon$ to analyse its statistics. Having said these generic properties of stochastic COP, we also notice that, $P(\epsilon)$ for different contact time ratios are unimodal and unlike $P(\eta)$ in underdamped case, the time asymmetry has no significant effect on $P(\epsilon)$.  

\subsubsection{Ovedamped dynamics}
\begin{figure}[H]
\begin{center}
\includegraphics[width=12cm]{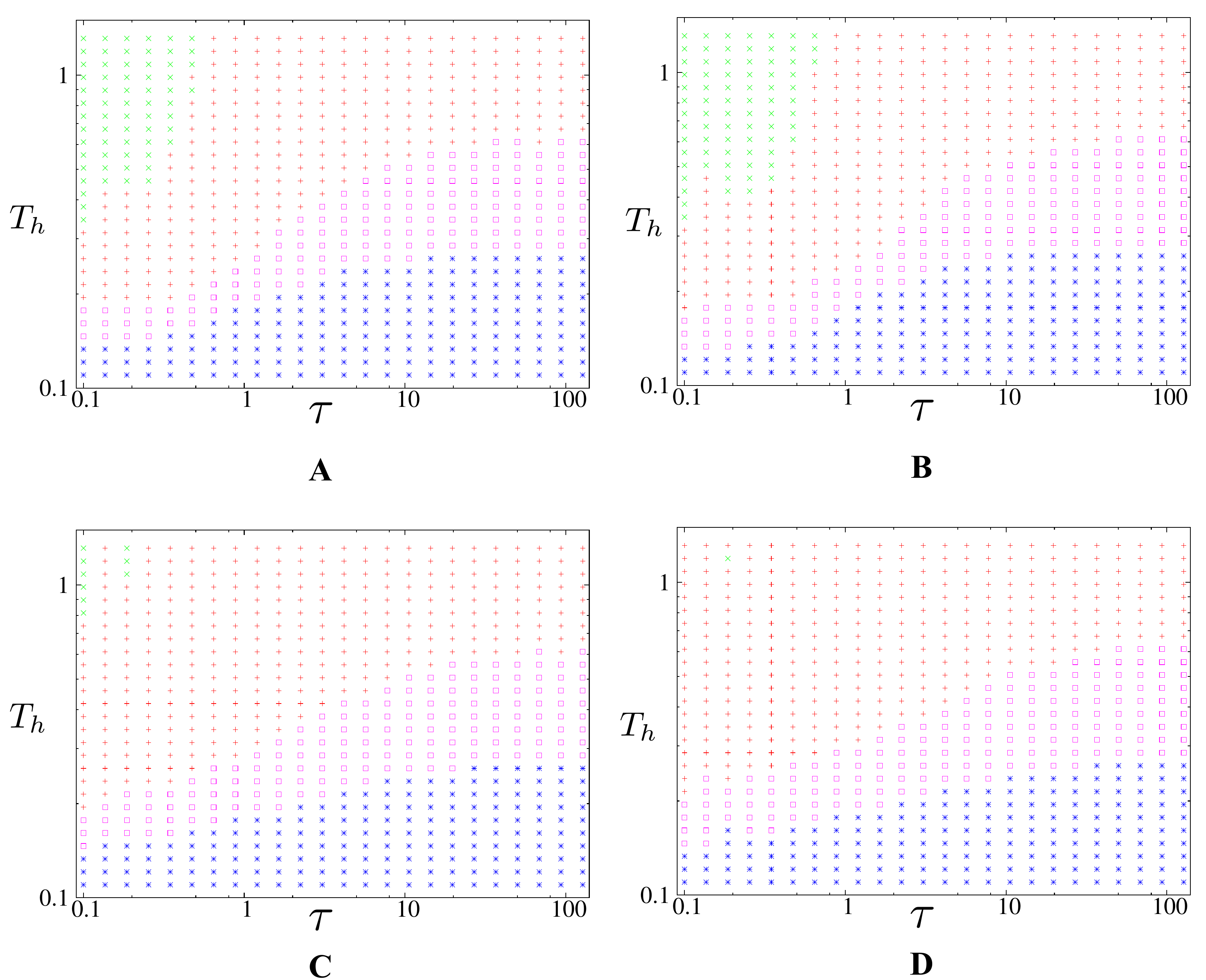}
\caption{ Different modes of operation of our system following overdamped Langevin dynamics with Carnot refrigerator protocol:  Open boxes(pink): heater-I, asterisk(blue): refrigerator, cross(green): engine, plus(red): heater-II. The ratio ($r:s$) of the contact time with hot bath during the isothermal compression to the contact time with the cold bath  during the isothermal expansion are:  A - 1:1, B - 1:3, C - 3:1 and D - 4:1.}
\label{ro}
\end{center}
\end{figure}

{\underline{Phase diagram}}: 
In Fig. \ref{ro}, we have plotted the phase diagrams for overdamped case with different time asymmetric protocols. Unlike the underdamped case, here we do not need any critical cycle time for the operation in refrigerator mode.  The system acts as a refrigerator for higher temperature differences compared to the earlier case. Therefore, total phase space area of the refrigerator mode has increased in this limit. The phase boundaries in quasistatic limit are consistent with our analytical results. The area of the engine mode in overdamped regime can also be monitored by changing $\tau_1$ and $\tau_2$. From the figure, it is seen when $\tau_1<\tau_2$, the system works as an engine for large number of values of $\tau$ and $T_h$. As one decreases the expansion time and increases the compression time, the engine mode gradually vanishes.

{\underline{COP distribution}}:  $P(\epsilon)$, the distribution of stochastic COP, is shown in  Fig. \ref{coeff_un} for $\tau=10$ for three different contact time ratios, where all the properties are similar to that of $P(\epsilon)$ in underdamped limit except the fact that here we notice a shoulder in the negative side of the distribution.  It shows power law tails with exponent $\sim 2$.


\begin{figure}[H]
\begin{center}
\includegraphics[width=7cm]{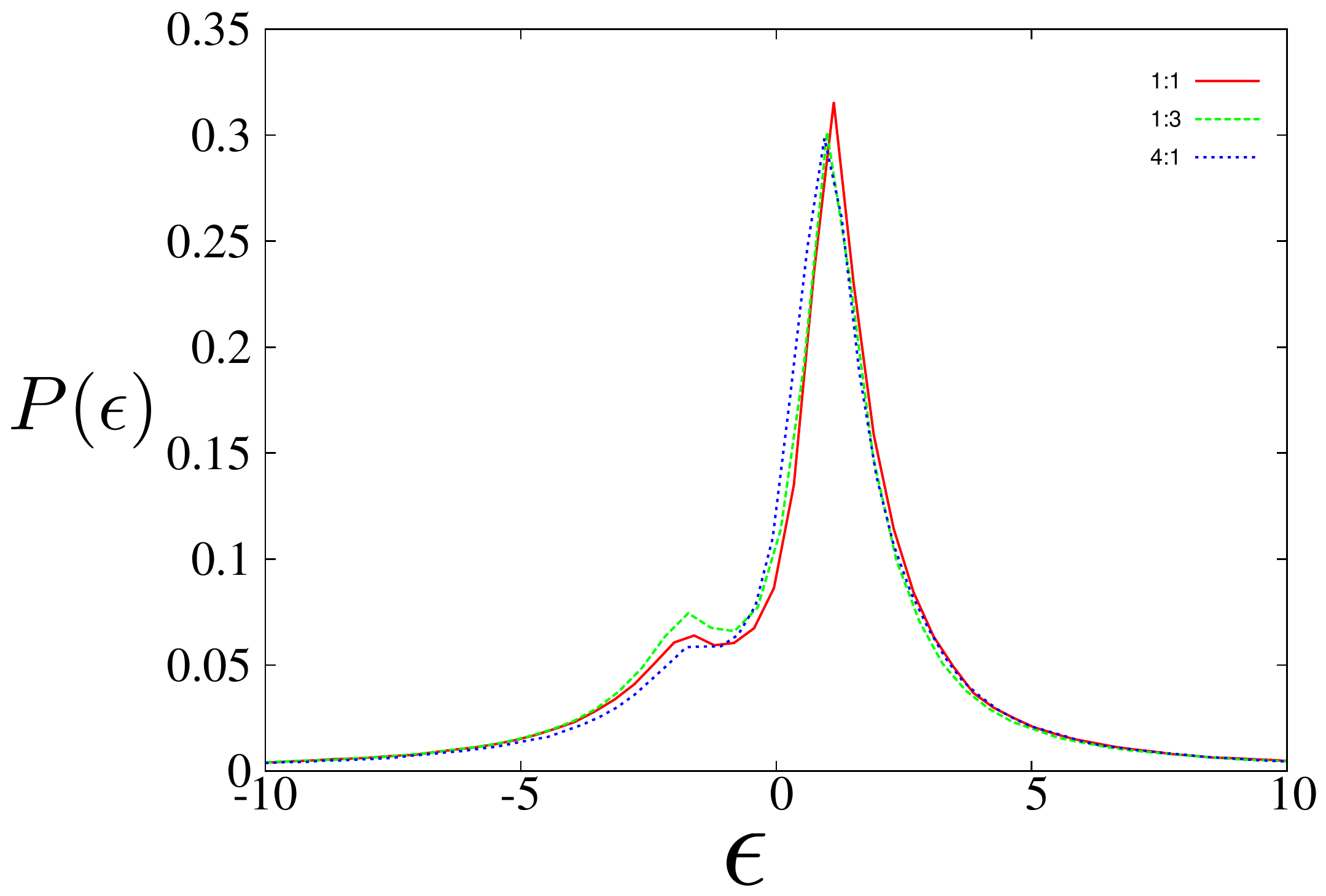}
\caption{ Distribution of COP for four different contact time ratios: red - 1:1, green - 1:3, blue - 4:1 at $\tau=10$ and $T_h=0.12$.}
\label{coeff_ov}
\end{center}
\end{figure}   

\section{Conclusion}
We have explored the operational characteristics and performance statistics of micro heat engines and refrigerators under time asymmetric protocols in detail. In our previous studies \cite{rana14,rana16} and also from the present work, it is evident that the microscopic engines are not equivalent to their macroscopic counterpart. Here, more importantly, we find that the time asymmetry of the protocol can considerably modify the phase diagram, depicting different operational modes of such micro-machines in $\tau-T_h$ plane. In non-quasistatic regime, we have shown that performance of such micro-machines depend on the time asymmetry of he protocol as it can affect the their performance statistics (e.g., $P(\eta)$ and $P(\epsilon)$). The distributions are broad and shows power law tails.  It has been shown that the power produced by the engine with asymmetric protocol is less than that of the engine with time-asymmetric protocol in underdamped limit. 


The jump heights of the protocol along the adiabatic steps play important role to determine whether the micro-heat engine or refrigerator can have reversible mode of operation in the quasistatic limit. Here, in high friction limit, a generic condition involving jump heights and bath temperatures is derived for which the micro-machine works in a reversible mode.

It is evident from the analysis that micro heat engines and refrigerators work differently under different protocols, particularly in non-quasistatic regime. We have also studied the thermodynamics of micro-machines driven by another important protocol, namely micro-adiabatic  protocol[]. We notice that in this case, the micro-machine can operate only in two different modes namely Heater-I and engine for different $\tau$ and $T_h$ and the phase diagram is comparatively simpler. Work along this direction is in progress.

\section{Acknowldgement}
A.M.J thanks DST, India for financial support and A.S thanks IOP, Bhubaneswar for local hospitality where the work has been carried out.

\end{document}